\documentclass[aps,twocolumn,epsfig,showpacs]{revtex4}
\usepackage{subfigure}
\usepackage{graphicx}
\usepackage{epsfig}
\usepackage{color}
\usepackage{amsmath}

\begin{document}

\title{On the rupture of DNA molecule}
\author{R. K. Mishra, T. Modi$^{1}$, D. Giri $^{1}$ and S. Kumar}
\affiliation{Department of Physics, Banaras Hindu University,
     Varanasi 221005, India \\
$^{1}$~Department of Physics, Indian Institute of Technology (BHU), Varanasi 221005, India}

\begin{abstract}
Using Langevin Dynamic simulations, we study effects of the shear force on the 
rupture of a double stranded DNA molecule. The model studied here contains 
two single diblock copolymers interacting with each other. The elastic constants 
of individual segments of the diblock copolymer are considered to be different.
We showed that the magnitude of the rupture force depends on whether  the 
force is applied  at $3'-3'-$ends or $5'-5'-$ends. 
Distributions of extension in hydrogen bonds and covalent bonds along the 
chain show the striking differences. Motivated by recent experiments, we have also 
calculated the variation of rupture force for different chain lengths. Results 
obtained from simulations have been validated with the analytical calculation 
based on the ladder model of DNA. 
\end{abstract}

\maketitle

\section{Introduction}

Separation of a double stranded DNA (dsDNA) is prerequisite for the essential 
cellular processes, such as, replication and transcription \cite{albert}. It is now well understood that DNA 
is stabilized by inter- and intra- molecular interactions \cite{albert,israel,kumarphys}. 
Single Molecular Force Spectroscopy (SMFS) 
techniques,  e.g., optical tweezers, magnetic tweezers and atomic force microscopy,  
have emerged as powerful tools to investigate these 
interactions \cite{kumarphys,Bockelmann, Bock,Lee_Science94, Strunge,Irina,prentiss1,hatch,Cludia,gaub,nik13}.
These experiments have provided various insights and understanding of biological 
processes at the molecular level. Moreover, experiments which explored the 
functioning of these interactions \cite{kumarphys,Bockelmann,Bock,Lee_Science94,Strunge,Irina,
prentiss1,hatch,Cludia,gaub,nik13} revealed that not only the magnitude of the force is 
important, but the nature of force, how and where force is applied, is also 
important in the understanding of the biological processes. To study DNA unzipping, 
a force has been applied perpendicular to the helix direction  \cite{Bockelmann, Bock}.
The unzipping force was found to be $\approx 15$ pN. In another experiment, the force is 
applied along the helical axis \cite{Lee_Science94}, and  rupture of DNA has been studied.
The rupture force is found to be significantly different than the unzipping force.
Strunz {\it et al.} \cite{Strunge,Irina} observed that the rupture force depends on the 
length of DNA and the loading rate.  The dynamics of dissociation of two strands was
also investigated \cite{neher}. Hatch {\it et al.} \cite{hatch} have performed systematic experiments on different lengths 
of DNA, and showed that the rupture force increases linearly for small chain lengths, but 
saturates at higher lengths.

To understand these observations, different  models of DNA have been 
proposed to explain the dynamics of DNA. Singh {\it et al.} \cite{singh} used a
lattice model of dsDNA, and showed that the rupture force increases with the length of DNA. 
Expressing the covalent and hydrogen bond as harmonic springs, 
de Gennes \cite{degennes} proposed that the net shear force required to rupture 
a homosequence dsDNA of length $l$ is  
\begin{equation}
f_c = 2 f_1 (\chi^{-1} \tanh(\chi \frac{l}{2})),
\end{equation}
where $f_1$ is the force required to break a single base-pair.
Here, $\chi^{-1}$ is the de Gennes characteristic length which is defined as 
\begin{equation}
 \chi^{-1} = \sqrt{\frac{Q}{2R}},
\end{equation}
where $Q$ and $R$ are  spring constants of covalent  and  hydrogen bonds respectively. 
Equation 1 shows that the shear force increases linearly with  $l$ (for small lengths), 
and saturates at higher values of $l$, which is consistent with the experiment \cite{hatch}. 
Different models similar to the ladder model have been studied and explored the different aspects of DNA 
rupture \cite{nelson,mishra, shikha, Tian}. 

In other experiments \cite{seol, ke},  the force-extension curve of a single strand DNA  consists 
of only Thymine (poly T) (or Urasil (poly U) in RNA) shows the 
entropic response, whereas Adenine (poly A)  shows  plateaus 
because of the base stacking.  As a result, the force-extension curves of these strands 
are found to be strikingly different \cite{seol, ke}. Thus, use of
different elastic constants for the complementary strands in the model studies is prerequisite.  Interestingly, theoretical 
models have used the same elastic constant for both strands in their description
\cite{degennes,nelson,mishra}.

Recently, Nath {\it et al.} \cite{nath} studied the DNA rupture using the ladder model, 
where the elastic constants of complementary strands were different. Their 
results were in good agreement with the atomistic simulations. However, 
because of the symmetry, the rupture forces applied 
at $3'-3'$ and $5'-5'-$ends will be the same. In contrast, Danilowicz  
{\it et al.} \cite{Cludia} have observed structural changes, when a force 
is applied at $3'-3'-$ends and $5'-5'-$ends, and the rupture force was found to be 
different. One of the possible reasons may be that the chain is heterosequence 
in nature, whereas theoretical models consider homosequence chain.

The simplest way to include  heterogeneity in the ladder model is to consider 
a single strand made up of a diblock  copolymer, i.e., half of the strand consists  
T type of nucleotides (say) and other half is of A type nucleotides (Fig. 1).  
In complementary strand, the first half consists of A type nucleotides, whereas 
second half is made up of T type nucleotides.
The elastic constant $Q$ of the segment consists of A is different than the 
elastic constant $U$ of the segment consists of T.
It is obvious from the Fig. 1 that the shearing force may be applied either at  
the ends having  nucleotides (A) of spring constants (Q-Q) (say, $3'-3'$) or  at the ends of  
nucleotides (T) having spring constants (U-U) ($5'-5'$) \cite{explanation}. 

The aim of this paper is to study the 
structural changes and to calculate the rupture force applied at the $3'-3'-$ and 
$5'-5'$-ends of dsDNA. In section II, we have developed a coarse grained model of DNA,  
and studied the dynamics of DNA under the shearing force using Langevin dynamic 
simulations. In section III, we showed that the rupture force  depends on the ends where 
the shearing force is applied.  In this section, we have also obtained the distributions 
of extension in  stretching of covalent and of hydrogen bonds, 
which give important information about the structural changes. 
Section IV contains the  analytical solution based on the ladder 
model of dsDNA consisting of a diblock sequence of DNA similar to the one developed 
in the Sec. II. The expressions for the  force required for rupture have been obtained 
for both cases (Q-Q and U-U), which are in good agreement 
with the simulation results.  The paper ends with a brief discussion on the results 
and future perspectives.

\begin{figure}[t]
\begin{center}
\includegraphics[width= 3.2in,height=1.2in]{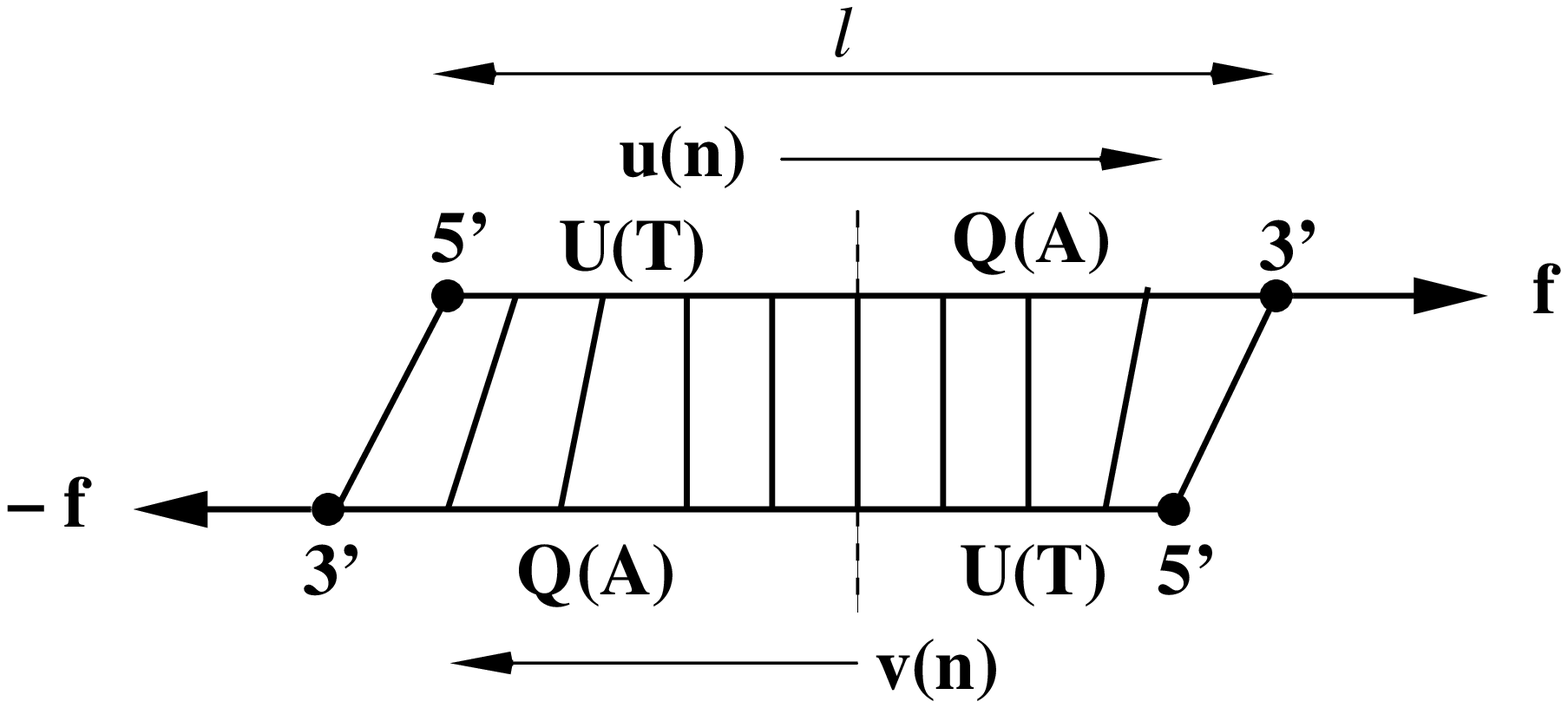}
\caption{Schematic diagram of a heterogeneous dsDNA preserving complementarity 
being pulled by a shear force.}
\label{fig1}
\end{center}
\end{figure}

\section{Coarse Grained Model of DNA}
A coarse grained model of two interacting flexible polymer chains in three dimension ($3d$) has been 
taken to model a dsDNA \cite{mishra}. The single strand of DNA  consists of two segments  connected by 
a covalent bond. In fact, each segment consists of nucleotide of one type called bead, which is composed  
of several molecules, e.g., sugar, phosphate, hydrogen, nitrogen etc. The interaction between 
consecutive beads (covalent bonds) is modelled by harmonic potential. 
The Leanard Jones (LJ) potential is used to model the base-pairing interaction 
between nucleotides of complementary strands. 
The total energy of the model system  can be 
expressed as: 

\begin{align}
E   & = \sum_{l=1}^2\sum_{j=1}^{N/2}k^{(l)}({\bf r}_{j+1,j}^{(l)}-d_0)^2
+ {\sum_{l=1}^2\sum_{j={N/2}}^N}k^{(l)}({\bf r}_{j+1,j}^{(l)}-d_0)^2 \nonumber \\
& +{\sum_{l=1}^2\sum_{i=1}^{N-2}\sum_{j>i+1}^N}4\left(\frac{C}{{{\bf r}_{i,j}^{(l)}}^{12}}\right) 
+ {\sum_{i=1}^N\sum_{j=1}^N}4\left(\frac{C}{(|{\bf r}_i^{(1)}-{\bf r}_j^{(2)})|^{12}} \right. \nonumber \\
& \left. - \frac{A}{(|{\bf r}_i^{(1)}-{\bf r}_j^{(2)}|)^6}\delta_{ij}\right),  
\end{align}
where $N$ is the number of beads in each strands. Here, ${\bf r}_i^{(\it l)}$ represents the 
position of the $i^{th}$ bead on the $l^{th}$ strand. In present case, $l = 1(2)$ corresponds 
to first (complementary) strand of  dsDNA. The distance between intra-strand 
beads, ${\bf r}_{i,j}^{(l)}$, is defined as $|{\bf  r}_i^{(l)}-{\bf r}_j^{(l)}|$. 
The first two terms of the above expression are the harmonic contributions from both
strands. In the first term, the spring constant  $k^{(1)} = U = 60$,  
whereas for the complementary strand $k^{(2)} = Q = 100 $ for the first half of the dsDNA,
i.e., $j= 1$ to $N/2$. The second term of above expression is the contribution 
of  harmonic terms for the remaining half of the strands ($k^{(1)} = Q = 100 $ and $k^{(2)} = U = 60 $). 
Third term takes care of the excluded volume effect, i.e., two beads
cannot occupy the same space \cite{book}. The fourth term  
corresponds to the LJ potential, which takes care of the mutual 
interaction between the two strands.
The  first term of LJ potential (same as third term of Eq. (3)) will not allow the 
overlap of two strands. Here, we set $C = 1$ and $A = 1$. The second term of LJ 
potential corresponds to the base-pairing interaction between two strands. The base-pairing 
interaction is restricted to the native contacts ($\delta_{ij} = 1$) only,  i.e.,  
the $i^{th}$ base of $1^{st}$ strand forms pair with the $i^{th}$ base of $2^{nd}$ 
strand only, which is similar to the Go model \cite{go}.
The parameter $d_0 (=1.12)$ corresponds to the equilibrium distance in 
harmonic potential. This is close to the equilibrium position of the LJ potential. 
The equation of motion is obtained from the following Langevin equation \cite{Allen,Smith,Li,MSLi_BJ07}:

\begin{figure}[t]
\includegraphics[width=3.2in]{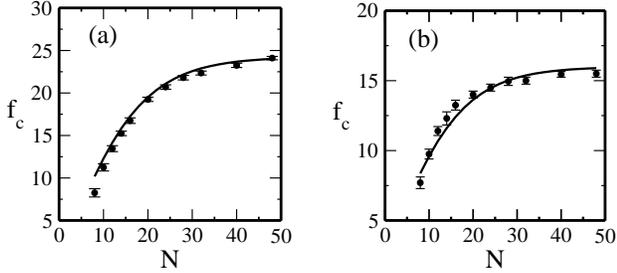}
\caption{Variation of rupture force with the length of dsDNA. The shearing force is
applied at (a) $3'-3'$ ends and (b) $5'-5'$ ends.
Solid circles (with error bars) are the data points obtained from the coarse grained simulations,
whereas the  solid line is obtained from the analytical expressions for the 
rupture force (Eqs. (24) and (25)).
}
\label{fig2}
\end{figure}

\begin{equation}
m\frac{d^2{\bf r}}{dt^2} = -{\zeta}\frac{d{\bf r}}{dt}+{\bf F_c}{(t)}+{{\bf \Gamma}(t)},
\end{equation}
where $m (= 1)$ is the mass of a bead and ${\zeta} (= 0.4)$ is the friction coefficient. 
The parameters used in Eqs. (3) and (4) are dimensionless.
Here, ${{\bf F_c}(t)}$  is given by ${-\frac{dE}{d{\bf r}}}$. The  random force ${\bf \Gamma}(t)$ 
is a white noise \cite{Smith},  i.e.,  $<  {\bf\Gamma} (t) {\bf \Gamma} (t')> 
= 2 d \zeta T \delta(t-t')$,  where, $d$ is the dimension of the space.
The choice of this dynamics keeps the temperature of the system constant
throughout the simulation. The equation of motion is solved by using the
6th order predictor-corrector algorithm with a time step of $\delta t = 0.025$ \cite{Smith}. 
In averaging,  different trajectories have been used and equilibrium
is assured by monitoring the stability of data with a longer run.

\section{DNA Rupture Analysis} 
We study  rupture events in the constant force ensemble \cite{kumarphys}. We  apply a constant force at the complementary 
ends ($3'-3'$ and $5'-5'$ ends) as shown in Fig. 1. and  add an energy $-\bf {f}.\bf {x}$ to the total 
energy of the system given by Eq. (3). 
Here, $\bf x$ is the elongation in DNA along the applied force direction. The rupture force is defined 
as a maximum force, when  all the  intact base-pairs break simultaneously. The most probable rupture force, 
obtained over many realizations, is referred as $f_c$ \cite{expl2}.
Fig. 2 (a)  shows the variation of rupture force with chain 
lengths, when the force is applied at the $3'-3'-$ends, whereas Fig. 2 (b) depicts the case when the force is applied at $5'-5'-$ends. 
It is evident from the plots that the magnitude of the force 
required to rupture dsDNA is  higher for the ends having  higher spring 
constant ($Q>U$). This is consistent with the hypothesis that higher force 
is required to stretch the stiffer bonds. It is also clear that the qualitative 
nature of the variation of rupture force with chain length is similar 
to the one seen in experiments,  i.e.,  for  small $N$, the rupture force 
increases linearly, and then saturates at higher values of $N$ \cite{hatch}.

\begin{figure}[t]
\includegraphics[width=3.2in]{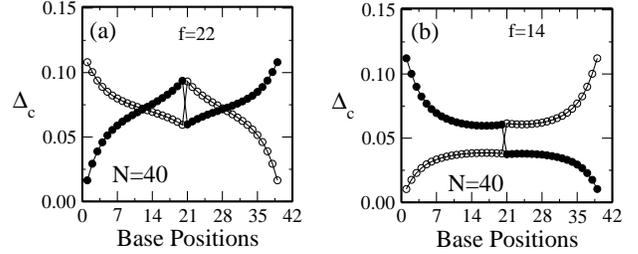}
\caption{Variation in the extension in covalent bonds along the chain.
A shearing force is applied  (a) at the $3'-3'-$ends and (b) at the $5'-5'-$ends. Solid and open circles 
correspond to the distribution of one (upper) strand and complementary (lower) strand, 
respectively (Fig.1).  The jump occurs at the interface of the segments of different elastic constants.}
\label{fig3}
\end{figure}

We now analyse the extension in covalent bond ($\Delta_{c}$) and stretching in 
hydrogen bond ($\Delta_{h}$) along the chain just below  the rupture force. Fig. 3 shows the variation of $\Delta_{c}$ 
with base positions. There is a striking difference between the distributions when the force is 
applied at $3'-3'-$ends  (Fig.3 (a)) and $5'-5'-$ends (Fig. 3(b)). It is clear that the distribution is symmetric 
like in previous study of homosequence of dsDNA \cite{mishra}, with a major change, i.e., occurrence of 
discontinuity  at the interface of segments of diblock model of dsDNA. When the force is applied at the 
$3'-3'-$ends, which has the larger elastic constant ($Q$), then the bonds near the pulling ends get stretched more, and decreases gradually. 
At the interface, because the net effect of force is transferring from higher elastic constant ($Q$) to the
lower elastic constant ($U$) side,  the extension in the bond increases at the 
interface ($f= - U x$), and there is a further decrease in the extension of bond 
towards the other end, i.e., $5'-5'-$ends. Here, the extension is quite less compare to the extension 
at the middle of the strand, because no force is applied at this end. However,
$5'-$ end has a base-pairing with $3'-$end, where a similar force is applied, but in the opposite direction. 
As a result, we see relatively less increase in the extension. Similar nature of the curve is obtained 
when the force is applied at the $5'-5'-$ends (Fig. 3(b)). The only apparent change is the decrease in the 
extension  at the interface of the segment of the diblock dsDNA. In this case the applied force is transferred 
from the lower elastic constant side (U) to the higher elastic constant side (Q).
Fig. 4 shows the distribution of stretching in 
hydrogen bonds $\Delta_{h}$ with the base positions. The characteristic de Gennes length for the present 
heterogeneous  system is about 10. It is clear 
that above this length, the  differential force approaches to zero. The discontinuity at the middle is attributed to the interface effect. 

\begin{figure}[t]
\includegraphics[width=3.2in]{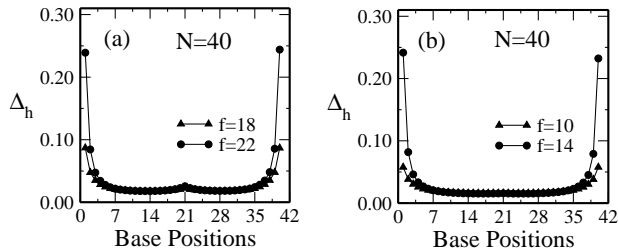}
\caption{Variation in extension of the hydrogen bonds $\Delta_{h}$ along the chain.  
The shearing force is applied (a) at the $3'-3'-$ends and (b) at the $5'-5'-$ ends}.
\label{fig4}
\end{figure}

\section{Ladder model of DNA}      

We revisit the ladder model of DNA
and include the heterogeneity in the description  to validate results obtained 
in Sec. III.  Following the de Gennes
formulation, we substitute the covalent bonds of adjacent nucleotides of each segment
with the harmonic springs. However, instead of taking uniform value of spring constant of each strand (Fig. 1), here, 
we choose spring constant ${U}$ for covalent bonds of the first segment of the single strand, 
and the spring constant ${Q}$ for covalent bonds of other segment of the same strand.   
For the complementary strand, $Q$ is the spring constant for covalent bonds of the first segment, while $U$ is the 
spring constant for covalent bonds of the other segment. 
The base-pairing interaction is also modelled by harmonic oscillator with the uniform 
spring constant $R$ assuming that interactions involved in  pairing are  the same for the  A-T and T-A base-pairs. 
Let the displacements of the upper and lower strands be  $u_{n}$ and $v_{n}$, 
respectively. The Hamiltonian of the composite diblock system can be written as

\begin{eqnarray}
H  &=& \sum_{n =0}^{\frac{N}{2}}\frac{1}{2} Q(u_n-u_{n+1})^2 + \sum_{n =-\frac{N}{2}}^{0}\frac{1}{2}U(u_n-u_{n+1})^2 \nonumber \\
&+& \sum_{n =0}^{\frac{N}{2}}\frac{1}{2} U(v_n-v_{n+1})^2 + \sum_{n =-\frac{N}{2}}^{0}\frac{1}{2}Q(v_n-v_{n+1})^2 \nonumber \\
 &+& \sum_{n =-\frac{ N}{2}}^{\frac{N}{2}}\frac{1}{2} R(v_n-u_n)^2
\end{eqnarray}

A shearing force may be applied either at $5'-5'$ or  $3'-3'-$ ends (Fig. 1). The first two terms of Eq. (5) correspond to 
the energy contribution due to the stretching of covalent bonds of the  upper strand. The next two terms 
are for the lower strand. The last term gives the energy contribution 
arising due to the stretching of the hydrogen bonds.

For $n > 0$, the equilibrium condition under the shear force for the upper strand can be written as

\begin{equation}
\frac{\partial H}{\partial u_n} \equiv Q(u_{n+1}-2u_n+u_{n-1}) - R(v_n-u_n) = 0, \label{eq:12}
\end{equation}
and similarly for the lower strand, it can be expressed as follows:

\begin{equation}
\frac{\partial H}{\partial v_n} \equiv U(v_{n+1}-2v_n+v_{n-1}) + R(v_n-u_n) = 0. \label{eq:13}
\end{equation}

\noindent If the total number of base-pairs is very large,  we can consider ${n}$ to be continuous, 
and thus, Eq. (6) and Eq. (7) can be written as

\begin{equation}
Q \frac{\partial^{2} u_{n}}{\partial n^{2}}-R(v_{n}-u_{n}) = 0  \label{eq:14}
\end{equation}
and 
\begin{equation}
U \frac{\partial^{2} v_{n}}{\partial n^{2}}+R(v_{n}-u_{n}) = 0.  \label{eq:15}
\end{equation}
From Eqs. (8) and (9), we obtained 

\begin{equation}
Q\frac{\partial^{2}u_n}{\partial n^2} + U\frac{\partial^{2}v_n}{\partial n^2}=0. \label{eq:16}
\end{equation}
Since ${u_n}$ and ${v_n}$ are independent from each other, therefore, Eq. (10) can be expressed as

\begin{equation}
\frac{\partial^{2}(Q u_n + U v_n)}{\partial n^2}=0.
\end{equation}
Integrating  Eq. (11), and the fact that the total tension is conserved, we obtained

\begin{equation}
Q u_n + U v_n = n f.  \label{eq:17}
\end{equation}
Dividing Eq. (8) by ${Q}$ and Eq. (9) by ${U}$, and subtracting, we get

\begin{eqnarray}
\frac{\partial^{2}(u_n-v_n)}{\partial n^2} - \frac{R(Q+U)}{QU}(u_n-v_n)  = 0,\\
\frac{\partial^{2}\delta_n}{\partial n^2} - \frac{R(Q+U)}{QU}\delta_n = 0,
\end{eqnarray}
where

\begin{equation}
\delta_n = u_n - v_n.   \label{eq:18}
\end{equation}

Equation (14) is the second order differential equation whose solution is of the form

\begin{equation}
\delta_n = \delta_0\cosh(\chi n)  + A\sinh(\chi n),    \label{eq:19}
\end{equation}
where  ${\chi^2 = \frac{R(Q+U)}{QU}}$. Here, $\delta_0$ is the elongation of the 
hydrogen bond at $n = 0$, and $A$ is an arbitrary constant of integration.
A similar solution exists for ${n < 0}$. Since the 
system has symmetry, the solution of Eq. (16) should also be symmetric and
therefore, $A$ will be zero. Using Eqs. (12), (15), and (16), expressions 
for ${u_n}$ and ${v_n}$ can be derived as

\begin{eqnarray}
u_n = \frac{n f}{Q+U} + \frac{U}{Q+U}\delta_0\cosh(\chi n) \\
v_n = \frac{n f}{Q+U} - \frac{Q}{Q+U}\delta_0\cosh(\chi n).
\end{eqnarray}
Similar expressions for the other side of the chain,  i.e., ${n<0}$ can be derived as

\begin{eqnarray}
u_n = \frac{n f}{Q+U} + \frac{Q}{Q+U}\delta_0\cosh(\chi n) \\
v_n = \frac{n f}{Q+U} - \frac{U}{Q+U}\delta_0\cosh(\chi n)
\end{eqnarray}

Since  harmonic potentials are used to simulate the interaction between base-pairs, an
additional parameter is needed to provide the condition for the rupture of hydrogen 
bonds. Let ${f_1}$ be the maximum force, which hold a base-pair intact, and  beyond that 
it undergoes rupture. Thus,

\begin{equation}
R\vert u_{n} - v_{n} \vert \geq f_{1}. \label{eq:10}
\end{equation}
The forces at the end points of the system must be balanced, which gives the
expression for the force as  

\begin{equation}
f = Q(u_{\frac{N}{2}} - u_{\frac{N}{2}-1}) + R(v_{\frac{N}{2}}-u_{\frac{N}{2}}).
\end{equation}
Substituting the values for ${v_n}$ and ${u_n}$ in Eq. (22), we get 

\begin{equation}
f=\frac{R(Q+U)}{U}\cosh(\chi \frac{N}{2})\delta_{0}(1+\chi^{-1}\tanh(\chi \frac{N}{2})). \label{eq:22}
\end{equation}
Using Eq. (21) and (23), we get the expression for rupture force applied at $3'-3'-$ends

\begin{equation}
\frac{f_{c}}{f_{1}}=\frac{Q+U}{U}(1+\chi^{-1}\tanh(\chi \frac{N}{2})).
\label{eq:23}
\end{equation}

When a force is applied at the end of the complementary strand ($5'-5'-$ends), the rupture force would be

\begin{equation}
\frac{f_{c}}{f_{1}}=\frac{Q+U}{Q}(1+\chi^{-1}\tanh(\chi \frac{N}{2})).
\end{equation}

\section{Conclusions}
In this paper, we have performed Langevin dynamic simulations using 
a coarse grained diblock model of dsDNA to study rupture events.  
We showed that magnitude of the rupture force is different when the force is applied 
at $3'-3'$ and  $5'-5'-$ ends. We have studied the distributions of 
extension in hydrogen and covalent bonds, which show the symmetry with a jump at the 
interface of the two segments. This discontinuity occurs due to the change in 
the elastic constants at the interface.  For short chains, we find that the rupture force increases 
linearly, and saturates for longer chains for both cases.
This is consistent with the experiment and previous studies \cite{hatch, mishra}. The distribution of hydrogen 
bonds shows that the differential force penetrates up to the de Gennes 
length. The numerical results have been validated with the ladder model of diblock DNA. We have obtained the 
analytical expression for the rupture force for both cases. It is evident from Fig. 2 that simulations 
data represented by solid circles are in good agreement with the analytical results (Eqs. (24) and (25)) shown by the solid line.
The de Gennes length in this case found to be $\chi^{-1} = \sqrt{\frac{Q U}{R (Q+U)}}$, which is in good agreement with simulation results.

At this stage, we must point out that though the model ignores the semi-microscopic detail of dsDNA,
e.g., precise description of $3'-3'$ and $5'-5'-$ends, orientation and inclination of base-pairs, 
helical structure of dsDNA, heterogeneity in the sequence, etc., even if, it captures 
some essential physics of the rupture mechanism of DNA.  
It would be interesting to perform all-atom simulations, where above shortcomings of the model
can be avoided, and one can get a better description of rupture mechanism. 

\section{Acknowledgments}
We thank Garima Mishra and B. P. Mandal for the fruitful discussions on the subject.
Financial supports from the Department of Science and Technology, and Council of
Scientific and Industrial Research, New Delhi are gratefully acknowledged. The
generous computer support from IUAC New Delhi is also acknowledged.


\begin{thebibliography}{99}
\bibitem{explanation} Inclusion of precise description of $3'-3'$ or $5'-5'$ of real DNA is beyond the scope of the present coarse grained model.
\bibitem{expl2} We have averaged over 25 realizations. The force has been incremented at the step of 0.5 till rupture occurs. The standard deviation found here is less than the force interval 0.5 used in the simulation.
\end{thebibliography}
\end{document}